# Analog Time–Frequency Analysis and Frequency Measurement of Ultra-Wideband Microwave Signals via Dual-Comb Channelization

Taixia Shi,[a,b] Wentao Ma,[a] Fangzheng Zhang,[b] and Yang Chen[a,*]

[a] Shanghai Key Laboratory of Multidimensional Information Processing, School of Information and Electronic Engineering, East China Normal University, Shanghai 200241, China
[b] National Key Laboratory of Microwave Photonics, Nanjing University of Aeronautics and Astronautics, Nanjing 210016, China
[*]Correspondence to: ychen@ce.ecnu.edu.cn

**ABSTRACT**
Frequency-to-time mapping (FTTM)-based photonics-assisted spectrum sensing enables wideband spectrum monitoring for applications such as cognitive radio and electronic warfare. However, FTTM-based photonics-assisted spectrum sensing approaches involve trade-offs among analysis bandwidth, temporal resolution, and frequency resolution. Although channelization has been introduced to alleviate these trade-offs, the limited number of available channels in previously reported systems constrains further improvements in overall performance. Here, we propose a photonics-assisted dual-comb channelization architecture for broadband frequency measurement and time–frequency analysis. The system employs a stepped-frequency pump based on stimulated Brillouin scattering (SBS) to generate an optical filter comb and adopts a single-laser design for the pump, probe, and reference paths to enhance frequency stability while reducing system complexity. Experimental results show a 48-GHz analysis bandwidth in the 16-channel configuration and 52 GHz in the 20-channel configuration, together with microsecond-scale temporal resolution and a frequency resolution better than 80 MHz. The proposed architecture therefore offers a practical and reconfigurable solution for real-time wideband spectrum sensing.

## 1. Introduction

Rapid and broad spectrum sensing over wide spectral ranges are crucial for effective spectrum management and deployment in applications such as cognitive radio [1], [2] and electronic warfare [3], [4]. In these scenarios, the spectrum sensing system is expected to cover a wide bandwidth while maintaining high frequency resolution and real-time operation. However, conventional electronic solutions are constrained by the bandwidth of analog-to-digital converters (ADCs) and the associated digital processing burden, which makes real-time analysis over ultra-wide frequency ranges difficult to achieve.

Microwave photonic technology [5], [6], leveraging the inherent advantages of optics such as broad bandwidth, high frequency, and low transmission loss, offers a promising solution for microwave signal generation [7], [8], [9], transmission [10], and processing [11]. Photonics-assisted spectrum sensing methods enable broadband spectrum analysis using limited-bandwidth ADCs, thereby avoiding the degradation in real-time performance associated with broadband electronic acquisition and processing. Consequently, microwave photonic spectrum sensing techniques [12], [13] have attracted extensive research interest. By combining photonics-assisted and electronic-based spectrum sensing methods, it is possible to achieve rapid, broadband, high-resolution, and accurate spectrum sensing [13], [14], [15].

Photonics-assisted spectrum sensing schemes mainly map signal frequency information into other measurable domains, such as power [16], [17, [18], space [14], or time. Among these approaches, frequency-to-time mapping (FTTM) is particularly attractive because it can process not only single-tone signals, but also multi-tone and time-varying waveforms. FTTM-based methods are generally implemented using either dispersion-based approaches [15], [19], [20], [21], [22], [23], [24], [25], [26], [27] or frequency-sweeping-and-filtering-based approaches [14], [28], [29], [30], [31], [32], [33], [34], [35], [36], [37], [38], [39], [40].

Dispersion-based FTTM methods recently can achieve analysis bandwidths exceeding hundreds of GHz with nanosecond-scale temporal resolution [15], [26]. However, such high temporal resolution generally requires an electrical sampling bandwidth of tens of gigahertz, placing stringent demands on high-speed ADCs and subsequent digital processing. Moreover, these methods are often constrained by the available dispersion, and their frequency resolution is typically limited to several hundred MHz. By contrast, although frequency-sweeping-and-filtering-based FTTM methods generally provide a coarser temporal resolution, typically on the microsecond scale, they can operate with a much lower electrical ADC sampling rate while offering greater reconfigurability and improved frequency resolution.

Significant research has been dedicated to enhancing the accuracy, frequency resolution, temporal resolution, and analysis bandwidth of frequency-sweeping-and-filtering-based FTTM spectrum sensing methods. Early one-dimensional frequency measurement studies improved the measurement accuracy through calibration and smoothing [28], and enhanced the frequency resolution by narrowing the stimulated Brillouin scattering (SBS) gain bandwidth [29] or nonlinear fitting [30]. By drastically increasing the sweep rate to shorten the sweep period, these frequency measurement techniques can be extended to include the additional temporal dimension, enabling microsecond-scale time–frequency analysis [33]. To extend the analysis bandwidth, methods employing sweep sources with larger bandwidths [38], multiple frequency-sweeping optical sidebands [39], or frequency-shift-loop [39] have been explored. However, these methods often require costly electrical sweep sources with high sampling rates and large bandwidths. To avoid the need for such expensive components, alternative methods that directly generate frequency-swept optical signals via optical means [34] or utilize tunable optical filters to replace frequency-swept optical signals [35] have been introduced. However, these approaches often suffer from poor sweep linearity, which compromises measurement accuracy and system stability. Furthermore, prior research [36] revealed that for a specific chirp rate, an optimal optical filter bandwidth exists to achieve the best frequency resolution. This finding highlights the intrinsic trade-offs among analysis bandwidth, temporal resolution, and frequency resolution in conventional single-channel FTTM systems.

Channelization has recently emerged as an effective strategy for mitigating these trade-offs [14]. By dividing the overall analysis bandwidth into multiple parallel channels, the total bandwidth can be expanded without proportionally increasing the bandwidth requirement of the electrical frequency-swept signal. In principle, this enables a wide analysis bandwidth while preserving the temporal and frequency resolution within each channel. However, the channelized implementation reported in [14] still exhibits several practical limitations. Its channel count is insufficient for further extending the analysis bandwidth. In addition, the use of multiple optical combs increases hardware complexity, while maintaining a sufficiently flat SBS gain comb becomes increasingly challenging as the number of channels increases. Moreover, the use of multiple laser diodes is susceptible to relative frequency drift, which degrades measurement stability and accuracy.

To address these limitations, a photonics-assisted dual-comb channelization system for broadband frequency measurement and time–frequency analysis is proposed. The proposed architecture uses a stepped-frequency pump signal to generate an SBS gain comb with a scalable number of channels. Meanwhile, the pump, probe, and reference optical paths are derived from a single laser source, thereby suppressing inter-laser frequency drift and reducing system complexity. The proposed system is experimentally validated in 10-, 16-, and 20-channel configurations. The results showcase a proof-of-concept for frequency measurement and time–frequency analysis, featuring a sub-100-MHz frequency resolution, a measurement error less than ±5 MHz, and a capability for wideband operation extending up to 52 GHz. In addition, an inter-channel crosstalk mitigation method is demonstrated to improve the fidelity of the reconstructed time–frequency results. Overall, these results indicate that the proposed architecture provides a practical and scalable solution for real-time wideband spectrum sensing.

## 2. Principle and experimental setup

### 2.1 Principle

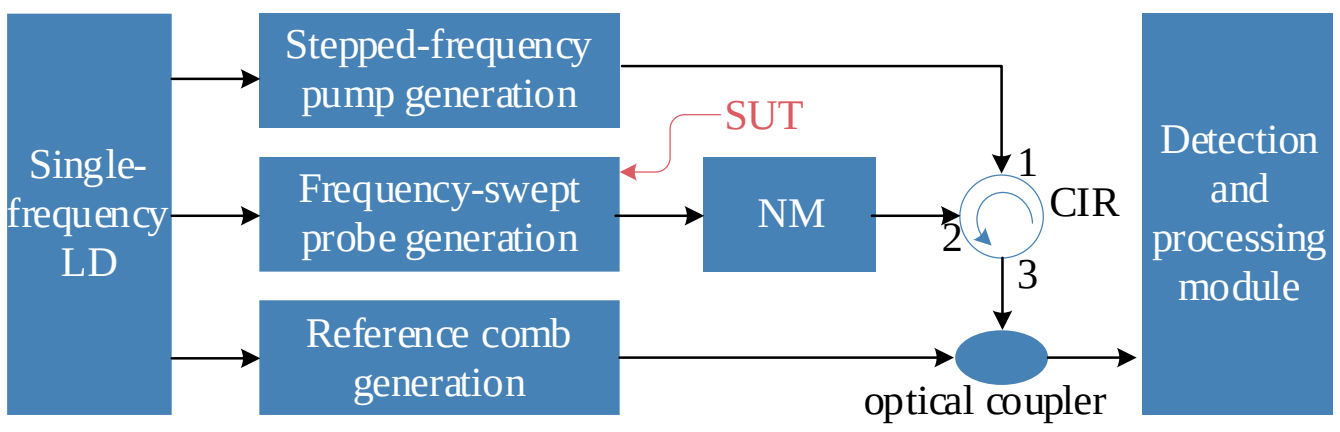

Fig. 1. Schematic diagram of the proposed scheme. LD, laser diode; SUT, signal under test; NM, nonlinear medium; CIR, optical circulator.

The proposed system performs frequency measurement and time–frequency analysis by combining frequency-swept optical probing, SBS based channelized filtering, reference comb assisted channel identification, and digital processing. A simplified schematic of the operating principle is shown in Fig. 1, while the detailed experimental implementation and the corresponding signal spectra are presented in Fig. 2.

As shown in Fig. 1, the optical carrier generated by a single-frequency laser diode (LD) is divided into three paths: the step frequency pump generation path, the frequency-swept probe generation path, and the reference comb generation path. In the probe path, the signal under test (SUT) is modulated onto a linearly frequency-swept optical carrier.

In the pump path, a step frequency optical pump is generated and subsequently launched into the nonlinear medium (NM), where it produces an SBS gain comb. The SBS gain comb provides the narrowband frequency-selective responses required for parallel FTTM across multiple analysis channels. When a frequency-swept optical component overlaps with the corresponding SBS gain line, it is selectively amplified and mapped into an optical pulse at a frequency-dependent temporal position, thereby implementing FTTM within that channel.

Since the output pulses from different SBS channels may overlap in the time domain, a optical reference comb is introduced for channel identification. After optical coupling and photodetection, pulses from different SBS channels beat with different reference-comb lines and are translated into electrical subcarriers at distinct frequencies. These subcarriers are then separated by digital bandpass filters. Finally, the envelope of each filtered subcarrier is extracted, and the obtained envelopes are rearranged according to the channel order to reconstruct the time–frequency distribution of the SUT.

In this architecture, the overall analysis bandwidth is extended by increasing the number of parallel SBS-filtering channels. By contrast, the temporal and frequency resolutions within each channel are mainly determined by the frequency-swept optical probe and the SBS gain bandwidth [36]. The detailed generation of the step frequency pump, frequency-swept probe, and reference comb is described in the following subsection with the aid of Fig. 2.

### 2.2 Experimental setup

The experimental setup of the proposed analog frequency measurement and time–frequency analysis system is shown in Fig. 2. The system mainly includes three parts: a probe link for converting the SUT to frequency-swept optical signals, a pump path for generating the SBS gain comb, and a reference-comb path for channelization and channel identification.

A continuous-wave optical carrier with a power of 16 dBm at 193.461 THz is generated by LD1 (ID Photonics CBDX1-1-C-H01-FA). The optical carrier is split into separate paths for pump generation, probe generation, and reference-comb generation by two optical couplers (OC1 and OC2). In the pump path, one output of OC2 is sent to the optical stepped-frequency generation module, which can be achieved by two methods. In Method 1, an intensity modulator (IM1, Fujitsu FTM7938EZ) is biased at the minimum transmission point (MITP) and driven by a stepped-frequency electrical signal generated from one channel (CH1) of the arbitrary waveform generator (AWG1, Keysight M8195A, 64 GSa/s). The electrical signal, featuring a peak-to-peak amplitude of 500 mV, is amplified by an electrical amplifier (EA1, Centellax OA4SMM3) before driving IM1. The stepped-frequency electrical signal in Method 1 has a start frequency of $f_{s11}$, a stop frequency of $f_{s12}$, and a frequency step of $B_1$. As illustrated in Fig. 2(a-i), carrier-suppressed double-sideband modulation produces two symmetrically distributed step-frequency optical sidebands.

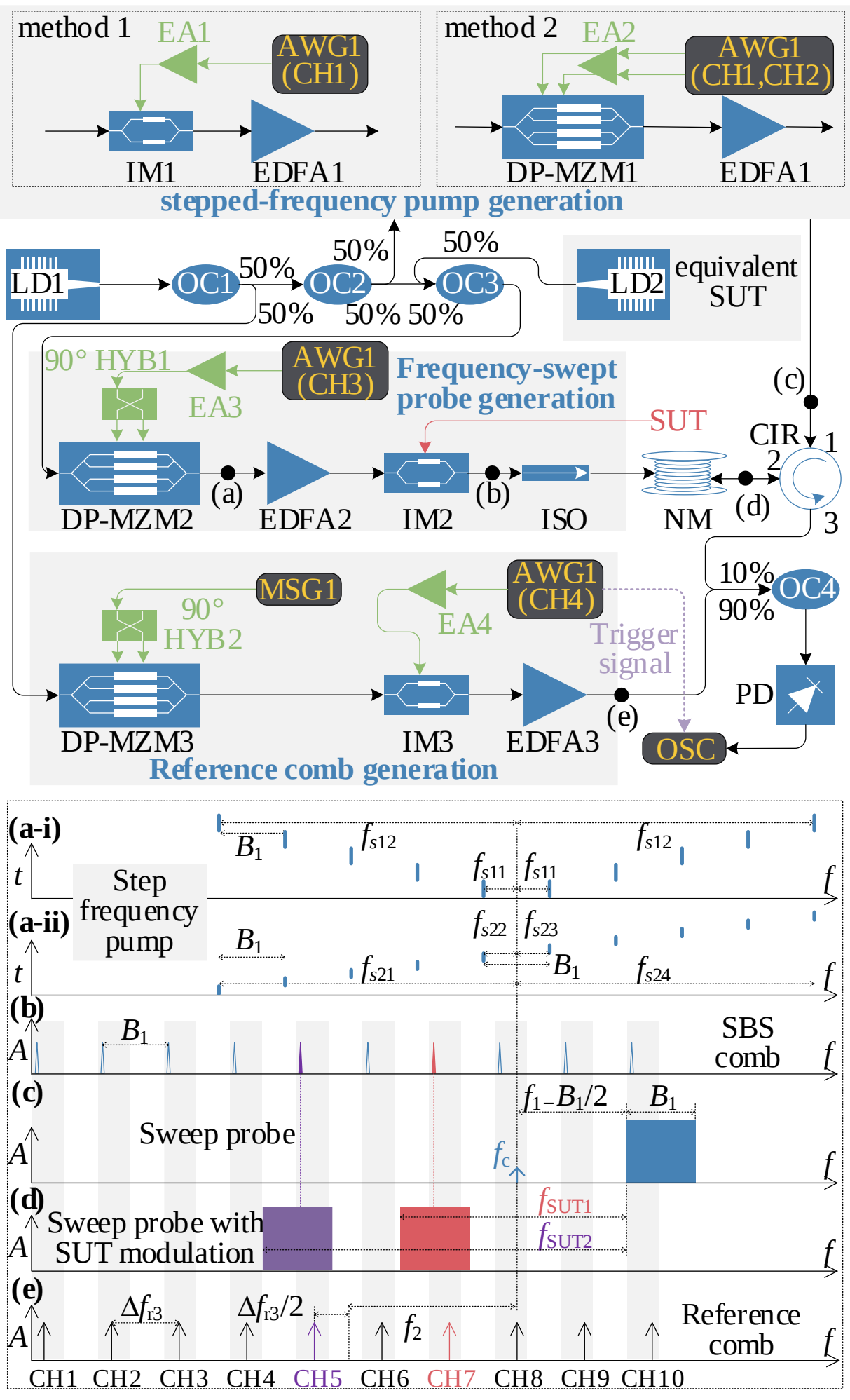

Fig. 2. Experimental setup of the proposed scheme. (a)–(e) Schematic of the signals at different positions in the system diagram. OC, optical coupler; IM, intensity modulator; AWG, arbitrary waveform generator; CH, channel; EA, electrical amplifier; EDFA, erbium-doped fiber amplifier; DP-MZM, dual-parallel Mach–Zehnder modulator; 90° HYB, 90° hybrid coupler; ISO, isolator; MSG, microwave signal generator; PD, photodetector; OSC, oscilloscope.

In Method 2, a dual-parallel Mach–Zehnder modulator (DP-MZM1, Fujitsu FTM7961EX) is biased as a carrier-suppressed single-sideband (CS-SSB) modulator and is driven by two V-shaped stepped-frequency electrical signals from CH1 and CH2 of AWG1 [15], each with a peak-to-peak amplitude of 300 mV. These two signals are amplified by a dual-channel electrical amplifier (EA2, Centellax OA4SMM4) before being applied to the DP-MZM1. They share the same frequency trajectory, while their relative phase flips with chirp polarity: +90° in the negative-chirp segment and −90° in the positive-chirp segment. The frequency of first steps from $f_{s21}$ to $f_{s22}$ with increment $-B_1$, and then steps from $f_{s23}$ to $f_{s24}$ with increment $B_1$, forming a V-shaped stepped-frequency trajectory. As shown in Fig. 2(a-ii), Method 2 generates a single-sideband step-frequency optical pump, avoiding the simultaneous symmetric sidebands produced by Method 1.

The optical signal from the optical stepped-frequency signal generation module is amplified by an erbium-doped fiber amplifier (EDFA1, Max-Ray EYDFA-C-HP-BA-33-SM-M), and then used as the pump wave and injected into the NM (a section of single-mode fiber with a length of 29.3 km) via an optical circulator (CIR) to generate the SBS gain comb. It should be noted that the SBS gain comb can be generated by a stepped-frequency pump since the period of the step optical signal is much less than the propagation time of the optical signal in the NM [41]. The resulting SBS gain comb is illustrated in Fig. 2(b). The spacing between adjacent SBS gain lines is designed to match the per-channel sweep bandwidth $B_1$, i.e., the analysis bandwidth of one channel, enabling contiguous channelization of the overall measurement range. For $N$ channels, the total analysis bandwidth is

$$B_A = NB_1. \tag{1}$$

If the lower boundary of the first channel is $f_{\mathrm{min}}$, the measurable frequency range is

$$f_{\mathrm{range}} \in [\, f_{\mathrm{min}}, f_{\mathrm{min}} + B_A \,]. \tag{2}$$

In the probe path, the optical carrier from OC2 is combined with an auxiliary optical source (LD2) only when equivalent wideband validation is required because of the limited frequency range of the available microwave sources. LD2 is not essential to the proposed architecture; with suitable high-frequency microwave sources, higher-frequency or wider-bandwidth SUTs can be measured directly without using LD2. The combined optical signal from OC3 is sent to DP-MZM2 (Fujitsu FTSM7961EX). DP-MZM2 is employed as a CS-SSB modulator. Its radio frequency (RF) ports are driven by a linearly frequency-modulated (LFM) signal generated by CH3 of AWG1, with a center frequency of $f_1$, a bandwidth of $B_1$, and a period of $T_1$. Before being applied to DP-MZM2, the LFM signal passes through EA3 (Centellax OA4SMM3) and a 90° hybrid coupler (90° HYB1, SHW TH-2/18-3S-90). The frequency-swept optical signal from DP-MZM2 is amplified by EDFA2 (Amonics AEDFAPA-35-B-FA), and then sent to IM2 (Fujitsu FTM7938EZ) which is biased at the MITP. As illustrated in Fig. 2(c), the frequency-swept optical probe covers a sweep bandwidth of $B_1$.

The SUT is loaded onto the frequency-swept optical signal at IM2. After modulation in IM2, the SUT with different frequencies is converted into frequency-swept optical signals corresponding to different frequency bands. As shown in Fig. 2(d), SUT components at $f_{\mathrm{SUT1}}$ and $f_{\mathrm{SUT2}}$ shift the frequency-swept optical components to different spectral positions. The optical signal from IM2 is injected into the NM via an optical isolator (ISO), and mapped to optical pulses in different channels by the SBS gain comb, and these optical pulses are output by port 3 of the CIR.

In the reference-comb path, another output of the OC1 is sent to DP-MZM3 for optical carrier frequency shifting. The RF ports of DP-MZM3 are driven by a single-tone signal with a power of 14 dBm and a frequency of $f_2$, which is generated by a microwave signal generator (MSG1, HP 83752B) and applied to DP-MZM2 via a 90° HYB2. The frequency-shifted optical signal from DP-MZM3 is sent to IM3 to generate the optical reference comb. IM3 is biased at the MITP, and its RF port is driven by an electrical comb with an output peak-to-peak amplitude of 500 mV, a frequency range of $f_{\mathrm{r1}}$ to $f_{\mathrm{r2}}$, and a frequency spacing of $\Delta f_{\mathrm{r3}}$. The optical comb signal is amplified by EDFA3 and then sent to the 90% input port of OC4. The 10% input port of OC4 is injected by the optical pulses from port 3 of the CIR. The optical reference comb is shown in Fig. 2(e). Its spacing is intentionally designed to be different from that of the SBS gain comb, so that the pulses from different SBS channels are converted into electrical subcarriers with distinct center frequencies after photodetection.

The combined optical signal from OC3 is sent to a photodector (PD). The optical pulses filtered by different SBS gain lines beat with different reference-comb lines, thereby generating electrical subcarriers with distinct center frequencies. Because $\Delta f_{r3}$ is set to be different from $\Delta f_{s1}$ or $\Delta f_{s2}$, each SBS channel is assigned a unique electrical frequency label, allowing the channelized signals to be separated by digital bandpass filtering. The electrical waveform is captured by an oscilloscope (OSC, R&S RTO2032 at 10 GSa/s and processed in MATLAB. Specifically, $N$ digital bandpass filters are used to separate the $N$ channels, the envelope of each filtered waveform is extracted, and the envelopes are rearranged to form the final time–frequency analysis result.

## 3. Experimental results

### 3.1 10-Channel Configuration

The proposed system in a 10-channel configuration is first experimentally verified and demonstrated. In this experiment, LD2 is turned off, and pump-generation Method 1 is employed. The relevant parameters are $f_1$=4 GHz, $B_1$=4 GHz, $T_1$=4 μs, $f_{s11}$=2 GHz, $f_{s12}$=18 GHz, $f_2$=10.2 GHz, $f_{r1}$=2.05 GHz, $f_{r2}$=18.45 GHz, $\Delta f_{r3}$=4.1 GHz. Here, $B_1$=4 GHz, represents the analysis bandwidth of each channel; thus, the 10 channels provide a total analysis bandwidth of 40 GHz.

A 0–6-GHz LFM signal with a peak-to-peak amplitude of 700 mV generated by AWG2 (Keysight M8190A) is first used as the SUT. The measured temporal waveform at the PD output is shown in Fig. 3(a), and the corresponding electrical spectrum is shown in Fig. 3(b). Based on the measured spectrum and the expected subcarrier positions, ten digital bandpass filters are designed with center frequencies ranging from 165 MHz to 1065 MHz, a spacing of 100 MHz, and a bandwidth of 20 MHz each. The spectra filtered by 10 different digital filters are simultaneously displayed in Fig. 3(c), and the waveforms corresponding to CH1–CH3 are shown in Fig. 3(d). It can be seen that the first two-thirds of the waveform appears in CH1, the last one-third appears in CH2, and no pulse is observed in CH3. This is because that the frequency range of the LFM SUT is 0–6 GHz and the frequency ranges of CH1–CH3 are 0–4 GHz, 4–8 GHz, and 8–12 GHz, respectively. In the zoomed-in view in Fig. 3(d), the pulses exhibit oscillatory variations because they contain both the pulse envelope and the high-frequency carrier component, rather than appearing as simple baseband pulses. To better visualize the pulses, the envelopes of the temporal waveforms in all channels are extracted, and Fig. 3(e) shows the envelopes of CH1–CH3. Based on the envelopes of the 10 channels, the time–frequency analysis result is obtained, as shown in Fig. 3(f), and Figs. 3(f-i) and 3(f-ii) show the zoomed-in views of CH1 and CH2 in Fig. 3(f), respectively. The LFM characteristic can be clearly observed from the time–frequency analysis result. For better visualization, the time–frequency analysis result presented in Fig. 3(f) is normalized, as indicated by the color bar. It should be noted that all subsequent time–frequency analysis results are processed using the same normalization method and presented with the same color-bar scale as that in Fig. 3(f).

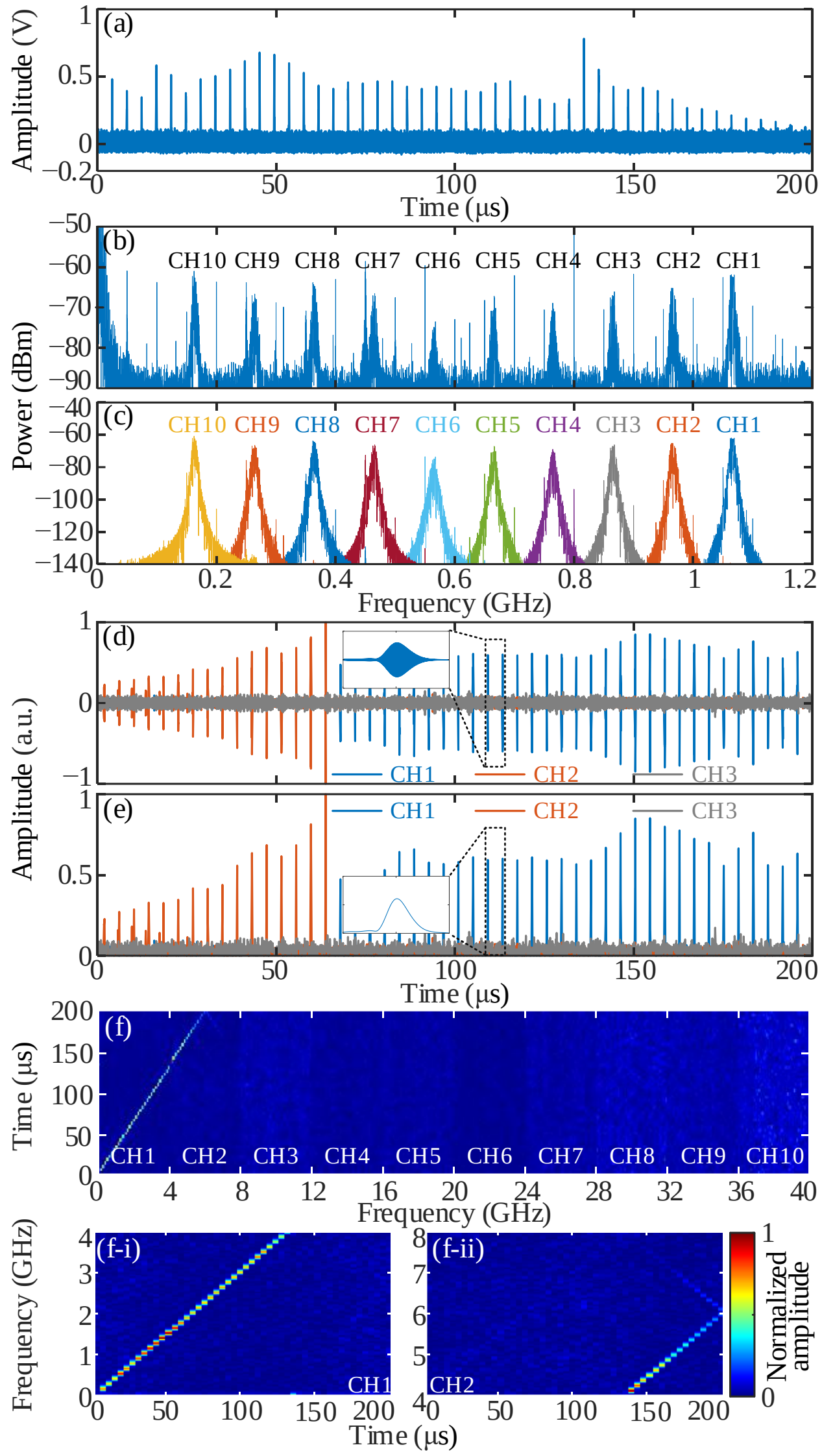

Fig. 3. Experiment results of 0–6 GHz LFM SUT. (a) Temporal waveform of the electrical signal from the PD. (b) Spectrum corresponding to (a). (c) Spectra of the signal after filtering by 10 digital bandpass filters corresponding to 10 channels. (d) Temporal waveforms of the signal after filtering by bandpass filters corresponding to CH1–CH3. (e) Envelope of the temporal waveforms in (d). (f) Time–frequency analysis results.

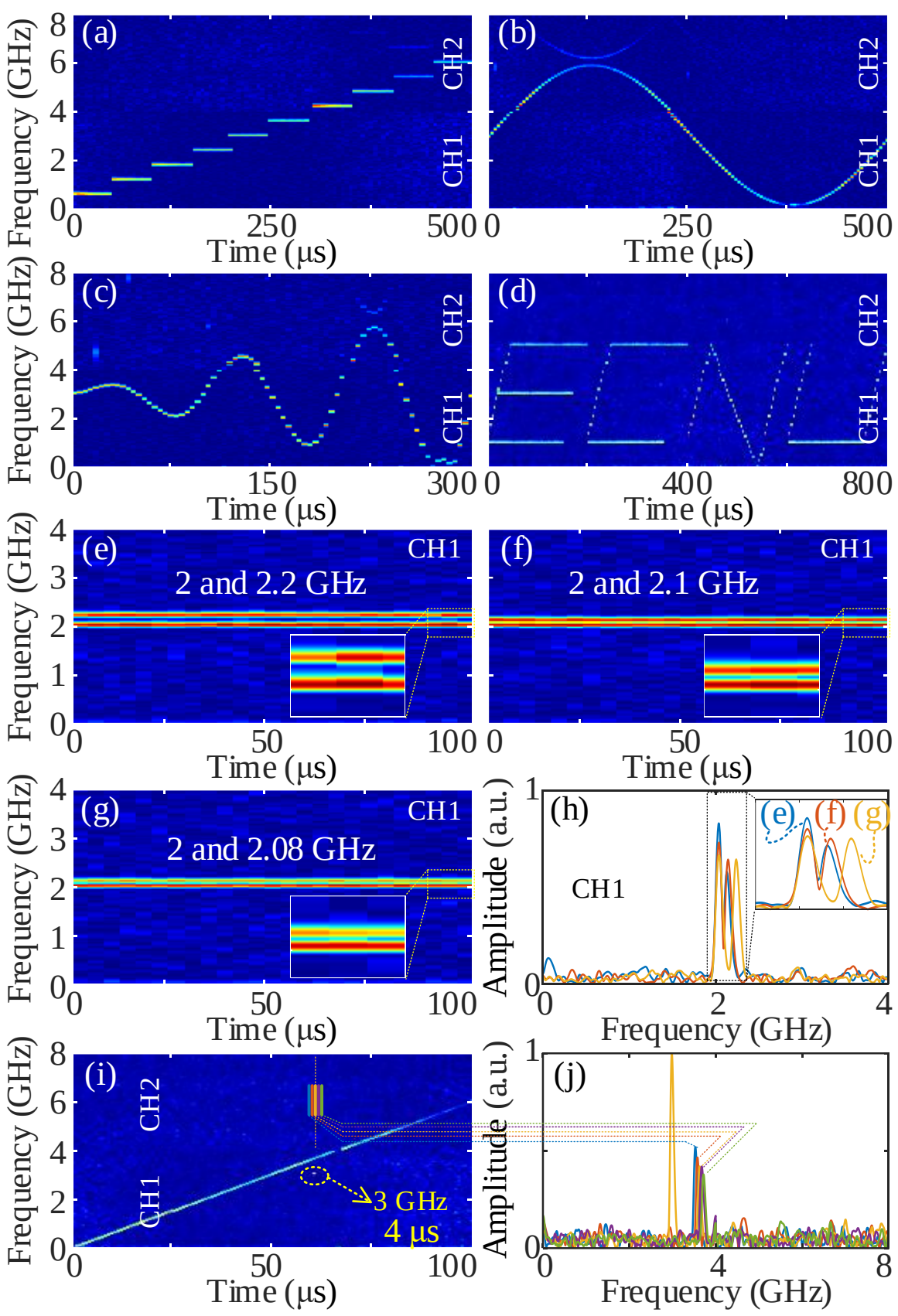

Fig. 4. Time–frequency analysis results of (a) step frequency, (b) signal with “Sine” time–frequency characteristic, (c) oscilating signal, (d) signal with “ECNU” time–frequency characteristic, (e) two-tone signal (2 and 2.2 GHz), (f) two-tone signal (2 and 2.1 GHz), (g) two-tone signal (2 and 2.08 GHz), (h) pulses waveform corresponding to (e)–(g) and (i) Time–frequency analysis results of an LFM signal and a burst signal. (j) Pulse waveform corresponding to (i).

To further demonstrate the capability of the proposed system in analyzing complex signals, various signal formats generated by AWG2 are further analyzed using the system. The corresponding time–frequency analysis results are presented in Fig. 4. These signals include: (a) a step-frequency signal, (b) a signal with a sinusoidal time–frequency characteristic, (c) an oscillatory frequency-modulated signal, (d) a signal with an “ECNU” time–frequency characteristic, (e)–(g) two-tone signals with frequency separations of 200 MHz (2 and 2.2 GHz), 100 MHz (2 and 2.1 GHz), and 80 MHz (2 and 2.08 GHz), respectively, and (i) a composite signal containing an LFM signal and a 3-GHz, 4-μs burst signal. The distinct time–frequency features of these specially designed signals are clearly resolved, as shown in Fig. 4. For instance, the two-tone signals in Figs. 4(e)–4(g) are successfully separated, demonstrating that the system provides a frequency resolution better than 80 MHz. The pulse waveforms corresponding to the two-tone signals in Figs. 4(e)–4(g) are shown in Fig. 4(h), which provide a direct view of the frequency resolution within a single measurement period. Furthermore, the pulse waveforms corresponding to the burst-signal case in Fig. 4(j), together with those from an adjacent measurement period, reveal additional details. When the burst signal is present, two distinct pulses are observed. In contrast, only a single pulse, corresponding to the continuous LFM component, appears in the adjacent measurement period. Additionally, due to the linear frequency-sweep characteristic of the LFM signal, the time position of the single pulse shifts linearly across successive measurement periods.

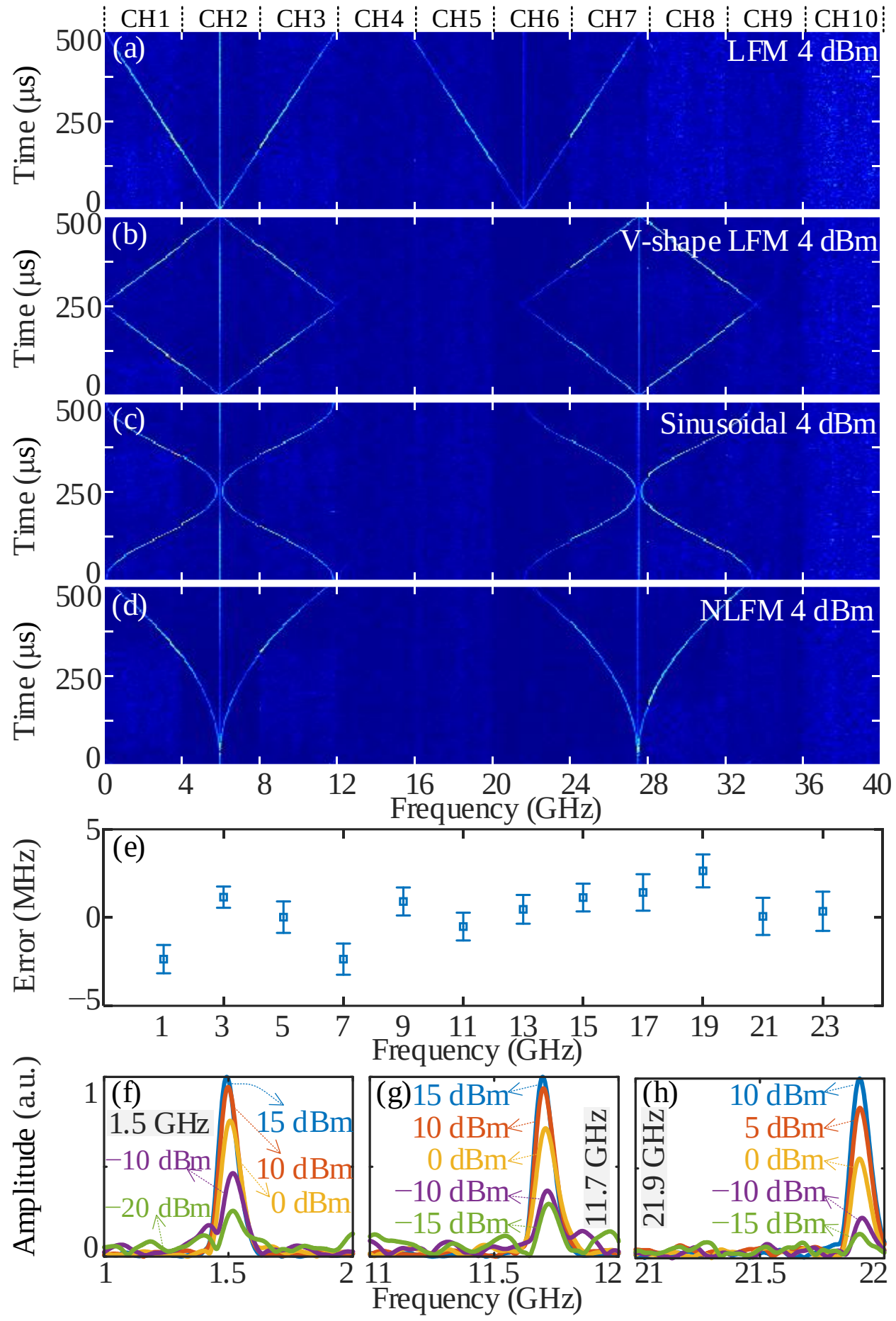

Fig. 5. Time–frequency analysis results of (a) LFM signal, (b) V-shape LFM signal, (c) signal with "Sine" time–frequency characteristic, (d) NLFM signal. (e) Frequency measurement error. (f)–(h) Measurement results at different SUT powers.

In order to evaluate the system's ability to analyze signals with larger bandwidth, LD2 is enabled. By tuning the frequencies of LD1 and LD2, their double-sideband spectra after SUT modulation can cover more channels, thereby equivalently demonstrating a larger analysis bandwidth. The time–frequency analysis results for several wideband signals generated by AWG2 are presented in Fig. 5: (a) an LFM signal, (b) a V-shaped LFM signal, (c) a signal with a sinusoidal time–frequency characteristic, and (d) an NLFM signal. The system accurately reconstructs the time–frequency trajectories of all these complex signals loaded onto the optical carriers corresponding to LD1 and LD2. The weak carrier-related component observed in Fig. 5 mainly results from incomplete optical-carrier suppression caused by DC-bias drift of IM2. The frequency measurement accuracy is quantified by measuring single-tone signals from 1 to 23 GHz in a step of 2 GHz, which are generated by MSG2 (Agilent 83630B) with its output power set to 9 dBm. As shown in Fig. 5(e), the measurement error remains within ±5 MHz. Finally, the system's detectable power range is investigated by measuring single-tone signals at 1.5 GHz, 11.7 GHz, and 21.9 GHz with varying input power. The pulse waveforms at different power levels are plotted in Fig. 5(f). The results indicate that the minimum measurable power is approximately −20 dBm at 1.5 GHz, −15 dBm at 11.7 GHz, and −10 dBm at 21.9 GHz. This frequency-dependent sensitivity is primarily caused by the degraded high-frequency response and increased loss of the modulators, amplifiers, cables, and other interconnects. Nonetheless, the system can detect signals over a wide frequency range. This frequency-dependent sensitivity is mainly caused by the degraded high-frequency response and increased loss of the modulators, amplifiers, cables, and other interconnects. Nonetheless, the system can detect signals over a wide frequency range.

3.2 16-Channel Configuration

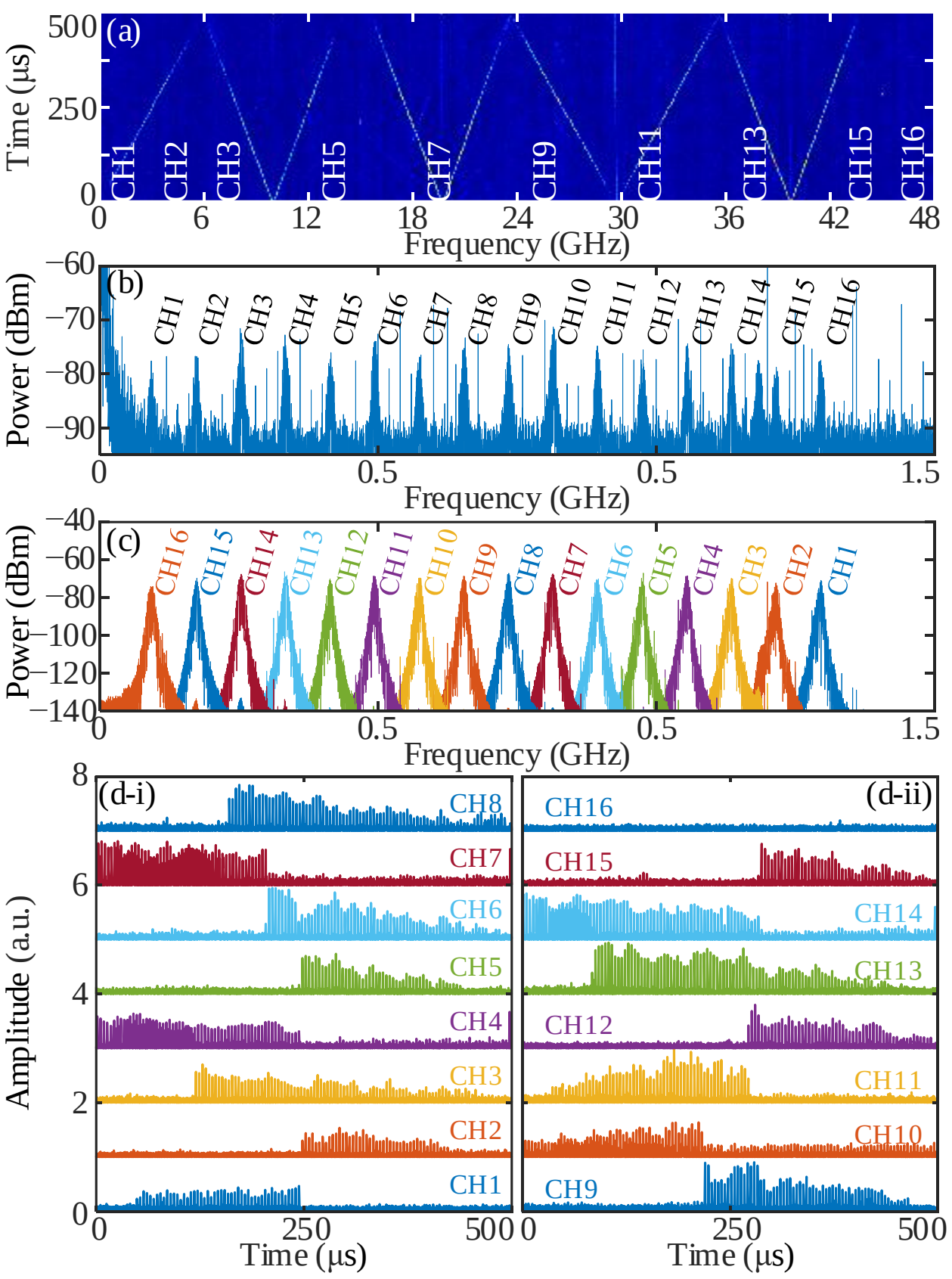

Fig. 6. Experiment results of LFM SUT. (a) Time–frequency analysis results. (b) Spectrum of the electrical signal from the PD. (c) Spectra of the signal after filtering by 16 digital bandpass filters corresponding to 16 channels. (d) Temporal waveforms of the signal after filtering by bandpass filters corresponding to CH1–CH16.

Then, the proposed system is demonstrated with the number of channels set to 16. In this study, the pump-generation Method 2 is employed. The relevant parameters are $f_1$=12.2 GHz, $B_1$=3 GHz, $f_{s21}$=2.5 GHz, $f_{s22}$=23.5 GHz, $f_{s23}$=0.5 GHz, $f_{s24}$=23.5 GHz, $f_2$=11.12 GHz, $f_{r1}$=1.54 GHz, $f_{r2}$=23.1 GHz, $\Delta f_{r3}$=3.08 GHz. Here, each channel has an analysis bandwidth of B1=3 GHz, and the 16 channels provide a total analysis bandwidth of 48 GHz. The center frequencies of the digital bandpass filters are set to the expected electrical subcarrier frequencies, ranging from 95 to 1295 MHz in 80-MHz increments. Each filter has a bandwidth of 20 MHz. The optical frequencies of LD1 and LD2 are set to 193.461 and 193.431 THz, respectively.

In this configuration, the frequency of LD2 is set to 193.431 THz. The SUT consists of two components. The first component, a 0–6 GHz signal from CH1 of AWG2, is fed into one input port of an electrical coupler. The second component, a 0–4 GHz signal from CH2 of AWG2, is upconverted to the 6–14 GHz frequency band using a mixer (Miteq M30) with a 10-GHz 10-dBm local oscillator (LO) signal supplied by MSG2 (Agilent 83630B). The upconverted signal is then fed into the second input port of the electrical coupler. The combined output of the electrical coupler, spanning 0–14 GHz, serves as the composite SUT.

The experimental results for a composite LFM SUT (0–6 GHz and 6–14 GHz) are presented in Fig. 6. The time–frequency analysis result in Fig. 6(a) clearly reconstructs the full linear sweep across the 14-GHz bandwidth. The corresponding spectrum of the electrical signal from the PD is shown in Fig. 6(b), while Fig. 6(c) displays the spectra of the signals after filtering by the 16 digital bandpass filters, corresponding to the 16 channels. The successful channelization is evident, with each filter capturing a distinct segment of the signal's bandwidth. The temporal waveforms of these filtered signals for all 16 channels are plotted in Fig. 6(d) over the frequency range from 0 to 48 GHz.

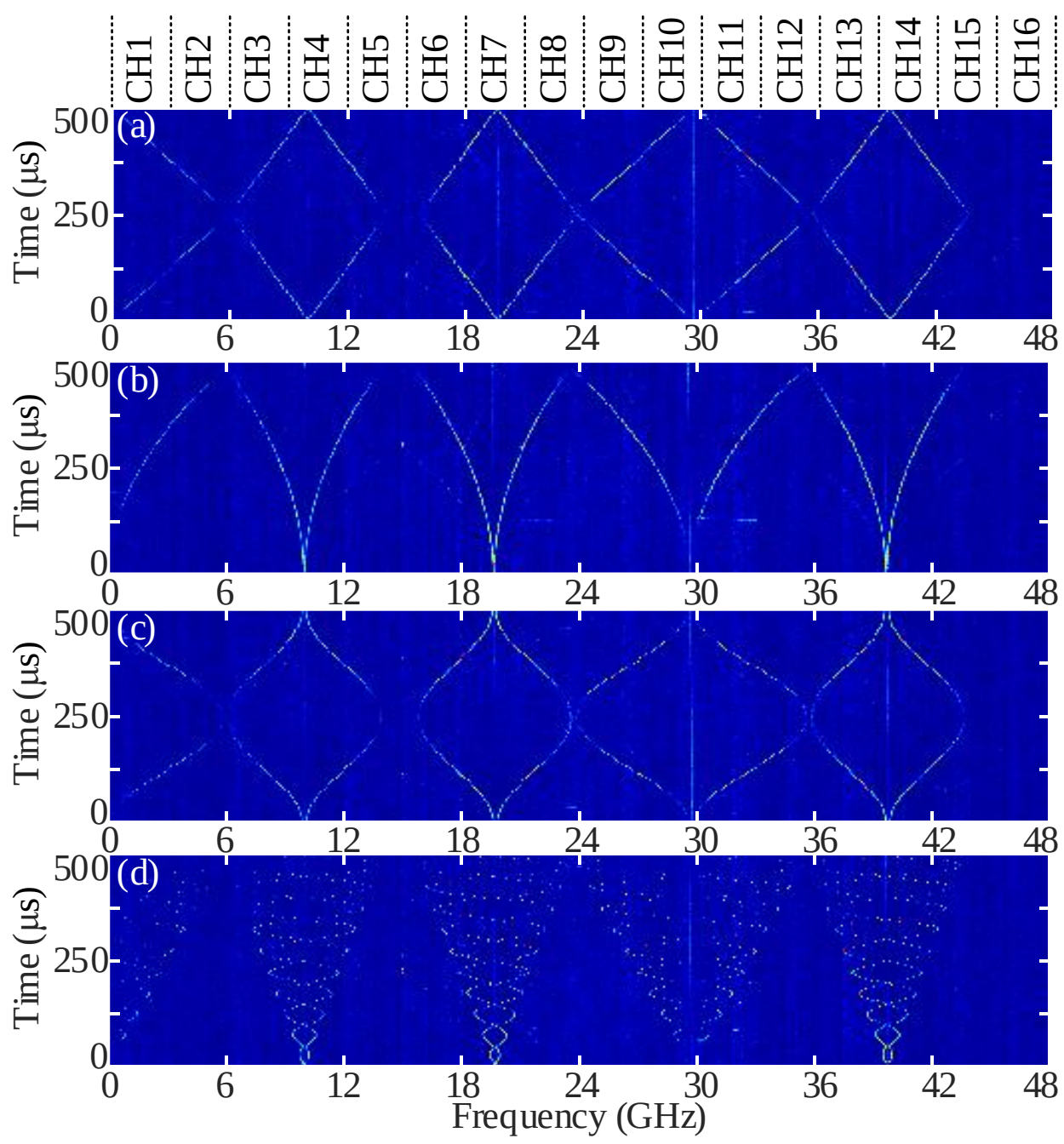

Fig. 7. Time–frequency analysis results of (a) V-shape LFM signal, (b) NLFM signal, (c) signal with "Sine" time–frequency characteristic, (d) oscillatory frequency-modulated signal.

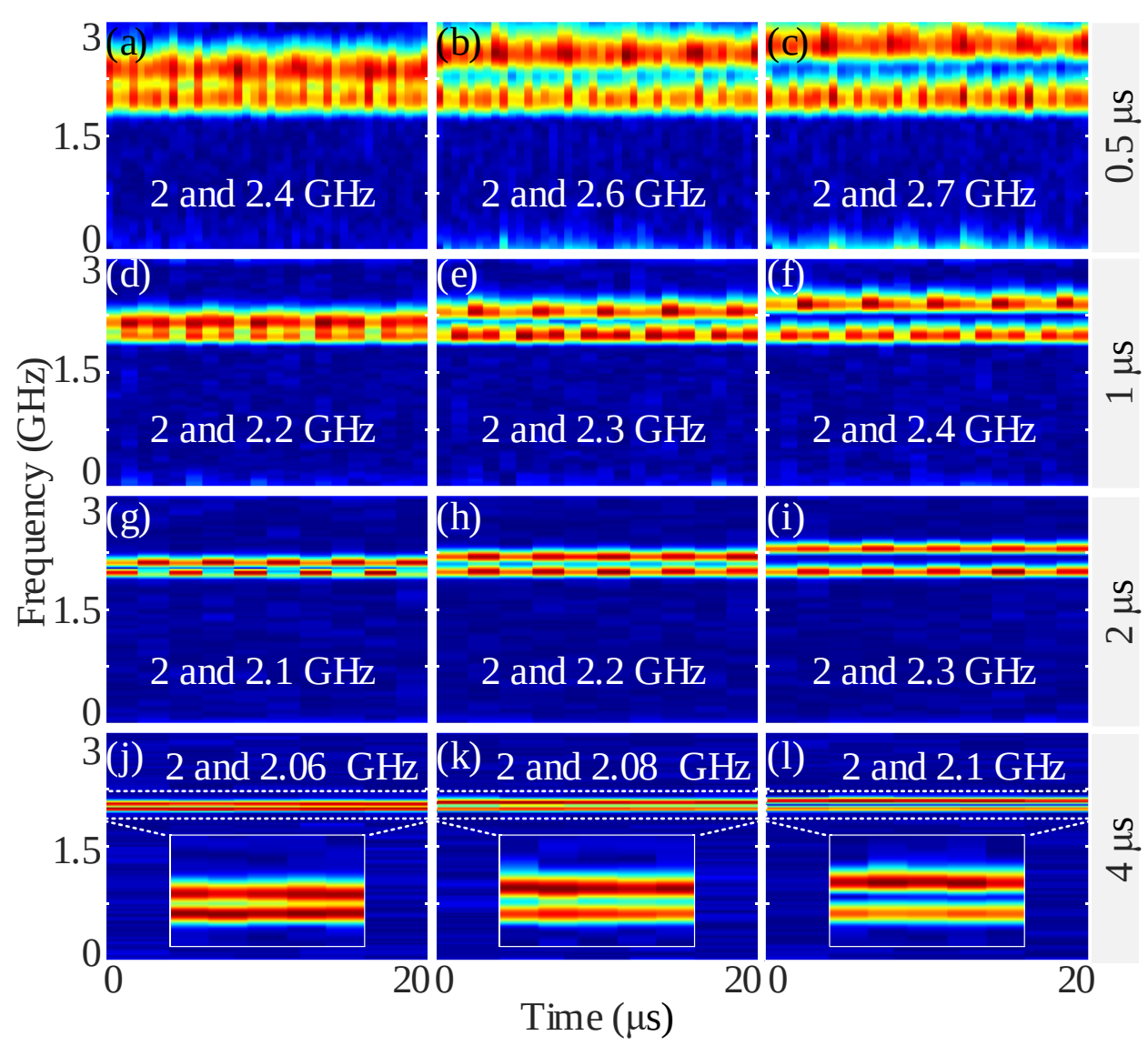

Fig. 8. Time–frequency analysis results at different two-tone signal frequency spaces and $T_1$.

Fig. 7 presents the time–frequency analysis results for the optical signals generated by modulating the frequency-swept optical probe with four different wideband waveforms under the same 16-channel configuration: (a) a V-shaped LFM signal, (b) an NLFM signal, (c) a signal with "Sine" time–frequency characteristic, and (d) an oscillatory frequency-modulated signal. The system accurately and simultaneously reconstructs the corresponding nonlinear time–frequency trajectories imparted by these modulation waveforms across the extended bandwidth, confirming its robustness for analyzing sophisticated modulation formats.

A further investigation into the system's frequency resolution is conducted using two-tone signals, with one tone fixed at 2 GHz and the other varied. The period $T_1$ of the LFM signal applied to DP-MZM2 is adjusted while keeping $B_1$ unchanged. The experimental results are shown in Fig. 8, establishing a direct relationship between $T_1$ and the achievable resolution. As shown in Fig. 8(a)-(c) with $T_1$ = 0.5 μs, two tones separated by 600 MHz are clearly resolved. Reducing the tone spacing requires a longer $T_1$: a 300 MHz separation is resolved with $T_1$ = 1 μs (Figs. 8(d)-(f)), a 100 MHz separation with $T_1$ = 2 μs (Figs. 8(g)-(i)), and finally, a 60 MHz separation with $T_1$ = 5 μs (Figs. 8(j)-(l)). This demonstrates that within each fixed-bandwidth channel, fine frequency resolution can be achieved through longer $T_1$, while the parallel channel architecture independently provides the wide total bandwidth.

### 3.3 20-Channel Configuration

Afterwards, the proposed system is demonstrated with the number of channels set to 20. In this study, the pump-generation Method 2 is employed. The relevant parameters are $f_1$=15.2 GHz, $B_1$=2.6 GHz, $f_{s21}$=1.3 GHz, $f_{s22}$=24.7 GHz, $f_{s23}$=1.3 GHz, $f_{s24}$=18 GHz, $f_2$=10.14 GHz, $f_{r1}$=1.33 GHz, $f_{r2}$=25.27 GHz, $\Delta f_{r3}$=2.66 GHz. Here, each channel has an analysis bandwidth of $B_1$=2.6 GHz, and the 20 channels provide a total analysis bandwidth of 52 GHz. The center frequencies of the digital bandpass filters are set to the expected electrical subcarrier frequencies, ranging from 105 to 1245 MHz in 60-MHz increments.

Each filter has a bandwidth of 20 MHz. The optical frequencies of LD1 and LD2 are set to 193.461 and 193.431 THz, respectively.

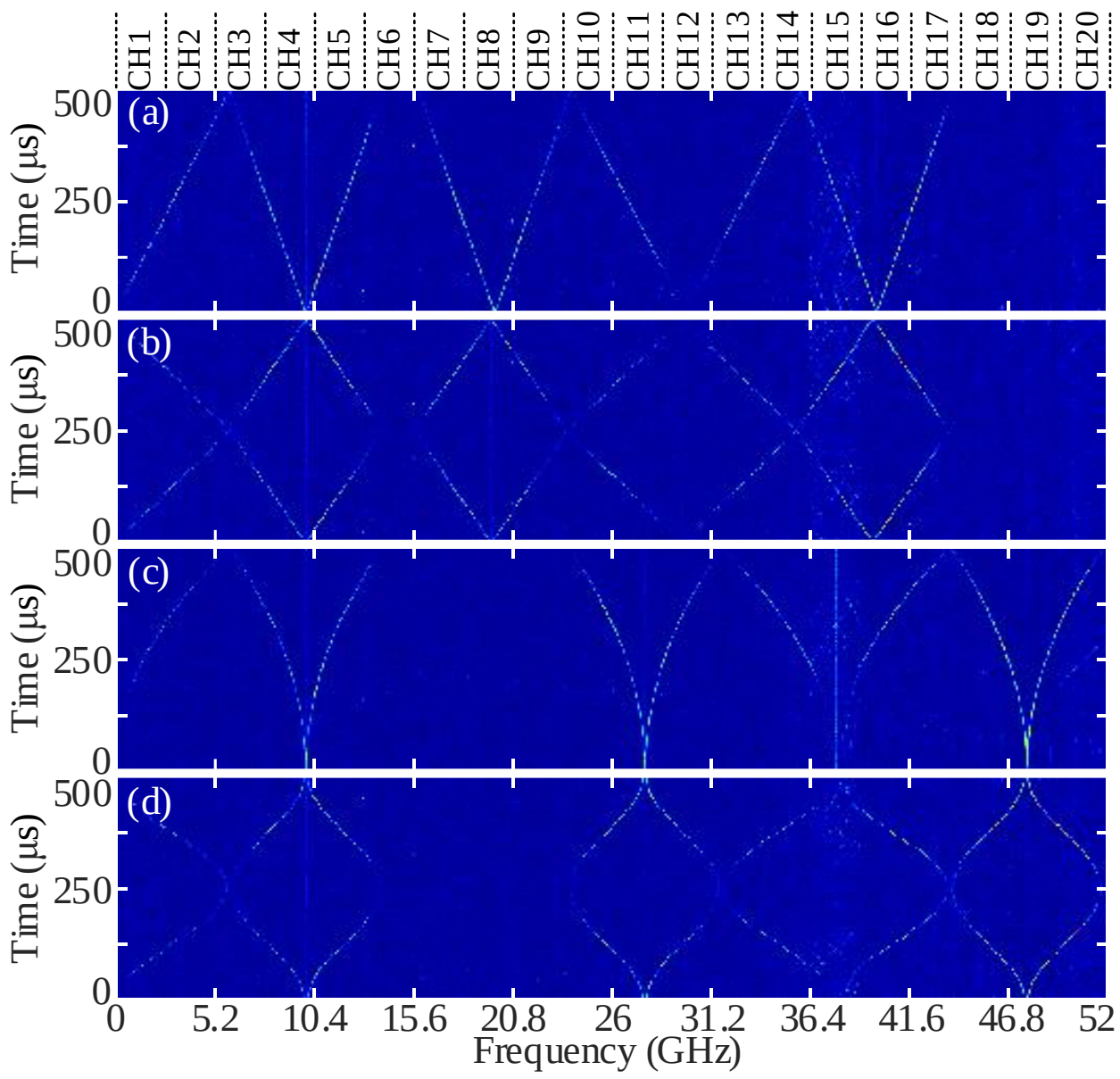

Fig. 9. Time–frequency analysis results of (a) LFM signal, (b) V-shape LFM signal, (c) NLFM signal, (d) signal with "Sine" time–frequency characteristic.

In this configuration, the analysis bandwidth is 52 GHz, and the composite SUT configuration in Fig. 6 is also employed. Fig. 9 presents the time–frequency analysis results for the signals generated by loading various wideband waveforms onto the frequency-swept optical probe under this ultrawideband configuration. The time–frequency trajectories imparted by four different modulation waveforms are accurately reconstructed: (a) a linear frequency modulation (LFM) signal, (b) a V-shaped LFM signal, (c) a nonlinear frequency modulation (NLFM) signal, and (d) a signal with a sinusoidal time–frequency characteristic. The analysis across the full 52-GHz span in a single measurement instance validates the system's capability for ultra-wideband signal characterization. It is noted that for optimal mapping of the signal spectrum onto the channelized comb, the frequency of LD2 is tuned to 193.431 THz for the measurements in Figs. 9(a) and 9(b), and to 193.423 THz for Figs. 9(c) and 9(d).

These results demonstrate that the proposed photonics-assisted dual-comb channelization approach is fundamentally scalable. By increasing the number of parallel channels, the total processing bandwidth can be significantly expanded without degrading the resolution of individual channels, a critical advantage over conventional single-channel wideband methods.

### 3.4 Removing the crosstalk between adjacent channels

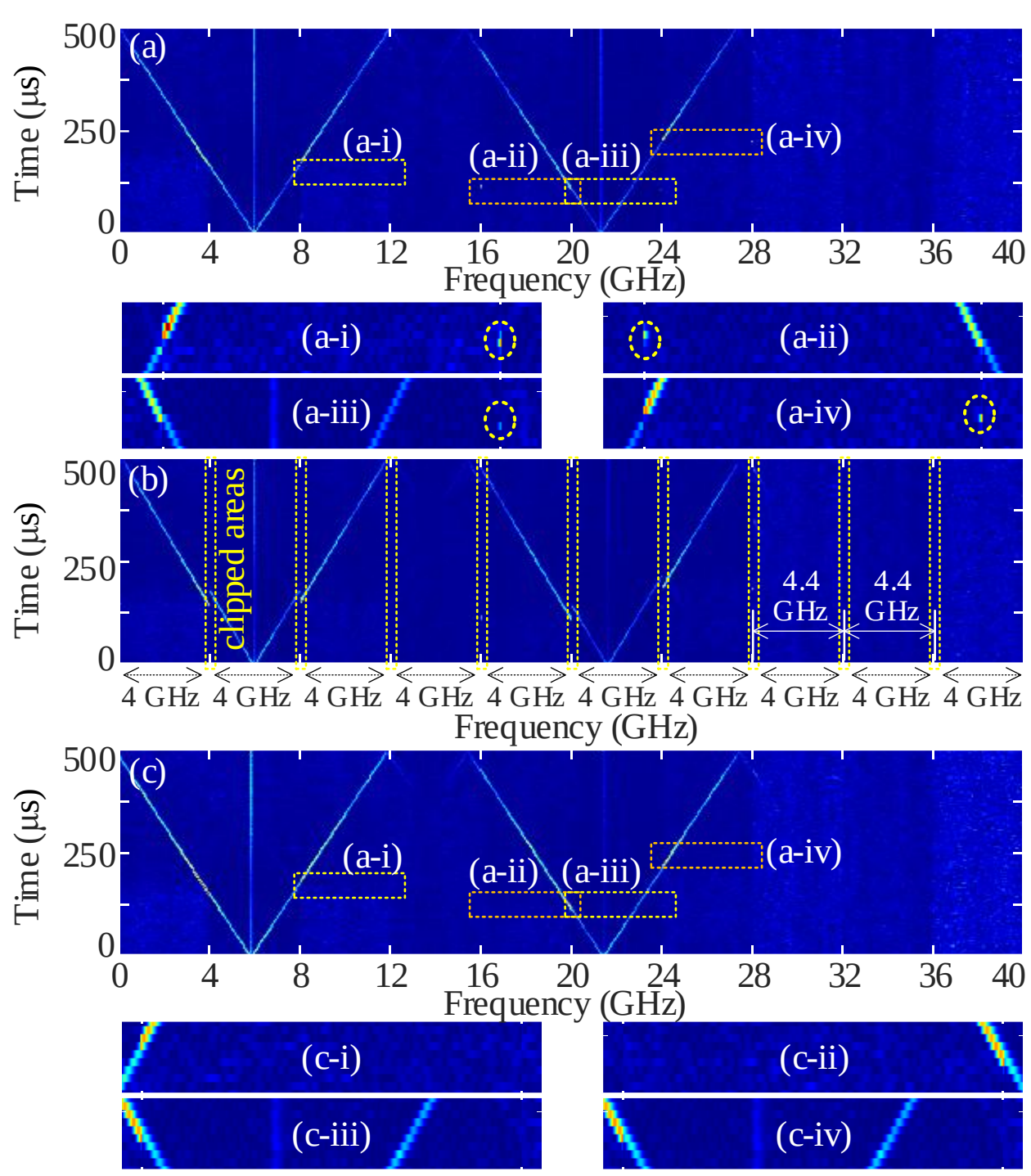

Fig. 10. (a) Time–frequency analysis results with crosstalk. (b) Time–frequency analysis results with broadened $B_1$. (c) Time–frequency analysis results after crosstalk removal.

A critical aspect in the design of multi-channel analysis systems is the management of inter-channel crosstalk. This study specifically investigated the crosstalk phenomenon inherent to the channelization architecture and proposed an effective mitigation strategy. The investigation is conducted under the 10-channel configuration, with initial parameters identical to those used in Fig. 3.

The manifestation of crosstalk is demonstrated in Fig. 10(a). When the frequency sweep bandwidth $B_1$ of the probing LFM signal is equal to the frequency spacing of the SBS gain comb, undesired crosstalk appears as the signal frequency approaches the boundary between adjacent channels. As detailed in the zoomed views of Figs. 10(a-i)–(a-iv), the temporal pulse corresponding to the signal has a finite width. When the instantaneous frequency is near a channel's boundary, a portion of this pulse's energy falls within the passband of the neighboring channel. This results in the signal incorrectly appearing in two adjacent channels simultaneously, creating a false time–frequency trace and degrading measurement fidelity.

To eliminate this crosstalk, a guard interval between channels is introduced. In this scheme, it is achieved by intentionally broadening the sweep bandwidth $B_1$, from 4.0 GHz to 4.4 GHz, without altering the spacing of the SBS gain comb or the optical reference comb. The increment (0.4 GHz) is configured based on the system's inherent frequency resolution. As shown in Fig. 10(b), this broadened $B_1$ ensures that the temporal pulse associated with a frequency at one channel's boundary completes fully before the instantaneous frequency enters the nominal passband of the next channel. Consequently, the crosstalk observed in Fig. 10(a) is eliminated. However, this operation creates an intentional spectral overlap in the analysis results, represented by the region within the yellow dashed boxes.

A final digital post-processing step is applied to obtain a clean, continuous time–frequency plot. The overlapped sections in Fig. 10(b) are cropped and spliced, yielding the result in Fig. 10(c). The zoomed views in Figs. 10(c-i)–10(c-iv) confirm the complete removal of the crosstalk artifacts, while the continuity of the signal's time–frequency trajectory is perfectly preserved. This method effectivelly suppress inter-channel crosstalk, thereby enhancing the accuracy and clarity of wideband signal analysis without requiring modifications to the hardware configuration.

## 4. Discussion

### 4.1 Pump signal generation and SBS gain flatness

In the proposed channelized system, each channel is required to have its independent SBS gain configuration. Therefore, generating multiple SBS gains simultaneously is a fundamental requirement. However, when multiple pump tones coexist, the available pump power must be distributed among them, while unequal tone powers and inter-tone nonlinear interactions may degrade the uniformity of the generated SBS gain comb. A critical challenge is maintaining adequate gain flatness across all channels. Poor SBS flatness leads to several issues: Signals may be undetectable in channels with insufficient gain, while channels with excessively high gain may suffer from an elevated noise floor, degrading the dynamic range and measurement consistency. This effect is illustrated in Fig. 11, which shows the measured temporal pulses for a single-tone signal when only one channel is active. With a fixed probe power, excessive pump power in that single channel introduces significant noise into the output pulse.

The challenge of achieving flatness in a multi-channel SBS gain comb becomes more pronounced as the number of channels increases. Fig. 12 presents the optical spectra of the pump signals and the corresponding equivalent SBS gain spectrum under different configurations. The pump spectrum is obtained by measuring the input optical signal of EDFA1, and the equivalent SBS gain spectrum is obtained at the output of port 3 of the CIR when no probe signal is injected. For a 10-channel system using pump-generation Method 1, a pump power variation of 1.35 dB across channels results in an SBS gain fluctuation of 4.1 dB, as shown in Fig. 12(a). When the channel count is increased to 16 while still using pump-generation Method 1, the pump power variation decreases slightly to 1.18 dB, but the SBS gain fluctuation worsens significantly to 8.64 dB, as shown in Fig. 12(b). A key limitation of pump-generation Method 1 is the generation of stronger spectral spurs, as two frequencies are present simultaneously, which contributes to the degraded SBS flatness at higher channel counts.

To facilitate the generation of a uniform SBS gain comb, a step-frequency pump is employed instead of a simultaneous multi-tone pump. For the 16-channel configuration, pump-generation Method 2 is used to generate a carrier-suppressed single-sideband step-frequency pump, with only one dominant optical frequency component present at each time step. This approach eliminates the spurious frequencies associated with concurrent multi-tone generation and, more importantly, allows for independent, time-multiplexed power adjustment for each pump tone, enabling precise power equalization across all channels.

As shown in Fig. 12(c) for a 16-channel system, the pump power variation is reduced to 0.69 dB, yielding a significantly improved SBS gain fluctuation of only 2.9 dB. For a 20-channel system, as shown in Fig. 12(d), the performance is further enhanced, with a pump power variation of 0.37 dB and an SBS gain fluctuation of 2.7 dB. Therefore, pump-generation Method 2 is essential for maintaining acceptable SBS flatness in systems with a large number of channels.

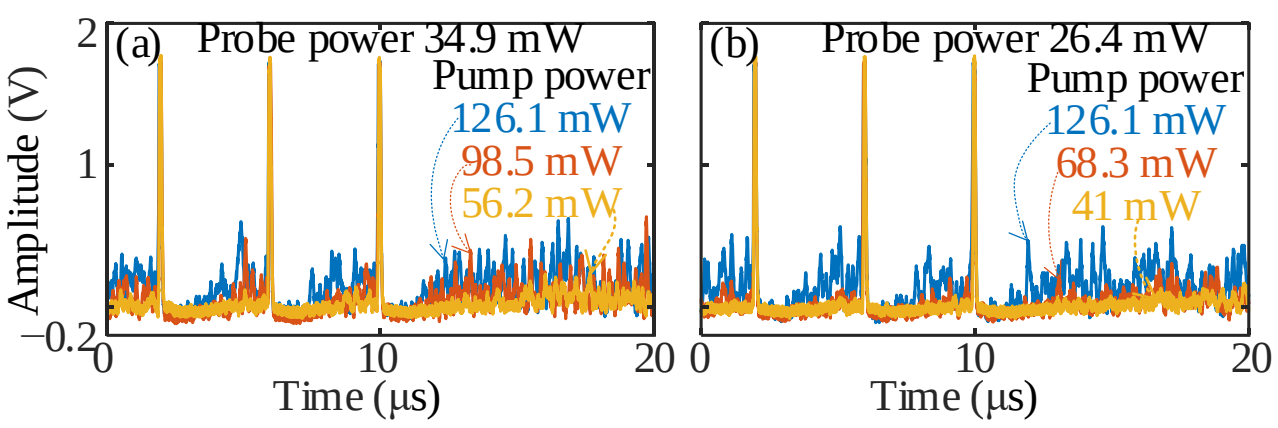

Fig. 11. Temporal pulse waveforms under different probe and pump power conditions.

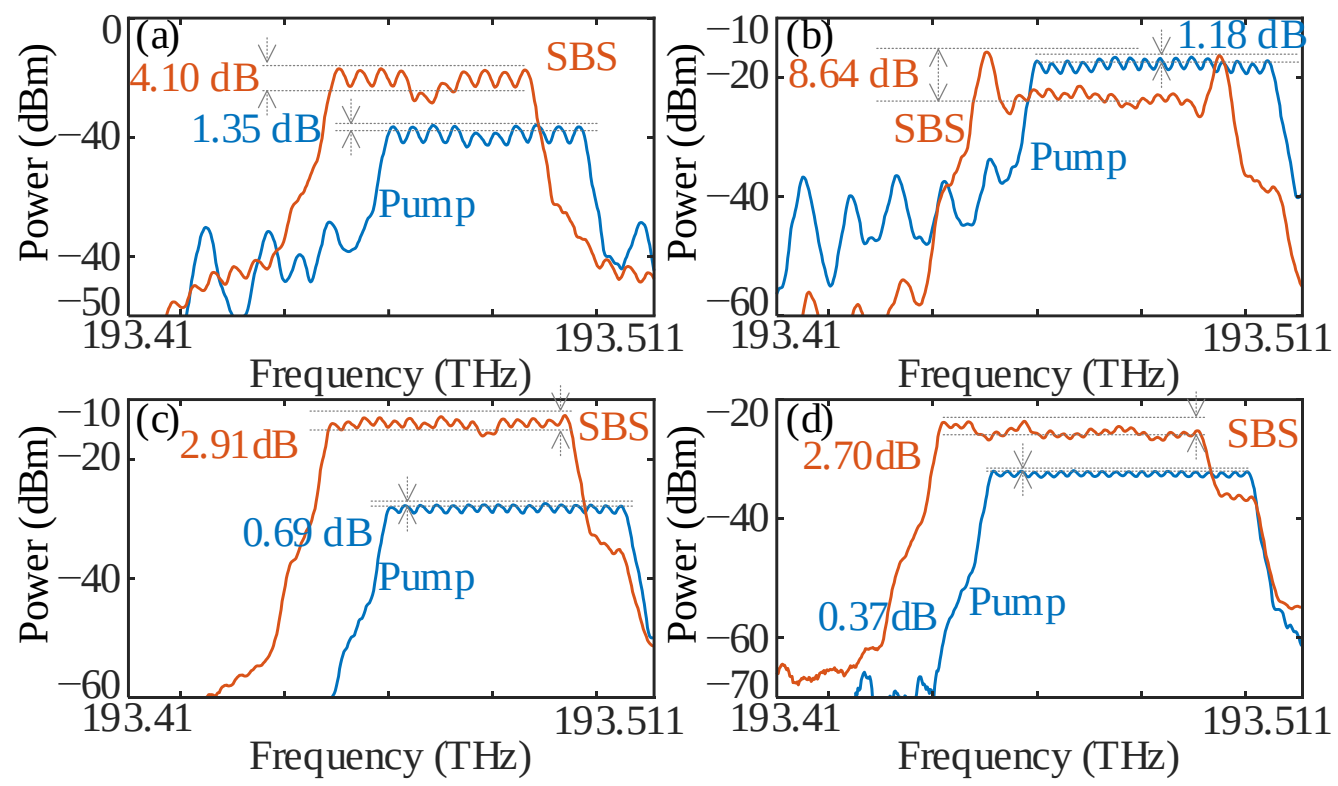

Fig. 12. Optical spectrum of pump and equivalent SBS when the channel number and pump generation method are (a) 10 and 1, (b) 16 and 1, (c) 16 and 2, (d) 20 and 2.

### 4.2 Alternative methods for pump and reference comb generation

In this experiment, the stepped-frequency pump signal and the optical reference comb are generated using a high-speed AWG. For practical implementation and reduced cost, alternative generation methods may be considered. The optical reference comb could be generated using cascaded electro-optic modulators driven by a single-frequency microwave source, which is a well-established approach for creating flat and broadband optical combs without relying on a high-sample-rate AWG [42]. Similarly, the stepped-frequency pump signal can also be generated using a recirculating frequency-shifting optical loop, where the optical carrier is shifted by a fixed frequency increment during successive round trips to form the required stepped-frequency sequence [43]. This approach offers an alternative to AWG-based generation for practical implementations.

### 4.3 Sampling rate and channel count limit

A key advantage of the proposed channelized architecture is that the channel count can be increased without duplicating the analog optical or electrical link. The SBS gain comb provides channelized narrowband optical filtering in a shared optical path, and the reference comb down-converts the filtered signals from different channels to distinct electrical subcarriers. The channelized signals are therefore frequency-division multiplexed in the electrical domain, with each subcarrier carrying the down-converted pulses from one channel. Consequently, the bandwidth that must be digitized is not the full analysis bandwidth, but rather the combined bandwidth occupied by the multiplexed electrical subcarriers. In the experiments, a sampling rate of 10 GSa/s is used for data acquisition, although it is not a fundamental requirement of the system. With the highest subcarrier centered at 1.245 GHz and a bandwidth of 20 MHz, a sampling rate of approximately 3 GSa/s would be sufficient in principle.

The system design inherently allows for the use of a lower sampling rate. As evident from the electrical spectra in Figs. 3(b) and 5(b), the total bandwidth occupied by the frequency-multiplexed subcarriers is less than half of the per-channel processing bandwidth $B_1$. Therefore, if $B_1$ is the bandwidth per channel, the required electrical sampling bandwidth is approximately $B_1/2$, enabling the use of a lower-speed and potentially lower-cost ADC. However, this introduces a fundamental constraint linking the maximum number of channels $N_{\max}$ to the subcarrier spacing $\Delta f_{sc}$. To prevent spectral overlap between adjacent subcarriers, the following condition must be satisfied:

$$N_{\max} \times \Delta f_{sc} \le \frac{B_1}{2}. \tag{2}$$

For instance, with $B_1$ = 4 GHz and a minimum feasible subcarrier spacing of $\Delta f_{sc}$ = 45 MHz (dictated by filter roll-off and necessary guard bands), the theoretical maximum channel count is $N_{\max} \leqslant$ (4 GHz / 2) / 45 MHz $\approx$ 44. This relationship highlights a design trade-off: utilizing a lower sampling rate to relax ADC requirements concurrently imposes an upper limit on the scalable number of channels for a fixed per-channel bandwidth.

The practical channel count is also limited by the available pump-power budget and the SBS interaction efficiency of the NM. Under fixed pump power and interaction conditions, increasing the number of channels reduces the effective pump resource associated with each SBS gain line, thereby lowering the per-line gain and the signal-to-noise ratio of the mapped pulses.

## 5. Conclusion

In summary, we have proposed and experimentally validated a photonics-assisted channelized architecture for wideband microwave frequency measurement and time–frequency analysis. The system effectively decouples analysis bandwidth from per-channel resolution by employing a scalable SBS gain comb and digital channelization, all implemented with a single laser source to enhance stability and reduce complexity. Experimental results demonstrate scalable performance up to 52 GHz analysis bandwidth with 20 channels, while maintaining 80 MHz frequency resolution and microsecond temporal resolution. The scheme operates with relatively moderate hardware requirements, highlighting its practicality for real-time, high-bandwidth spectrum sensing in fields such as cognitive radio, intelligent transportation, and electronic warfare.

**Acknowledgements**
National Natural Science Foundation of China (62371191, 62401207); Key Laboratory of Radar Imaging and Microwave Photonics (Nanjing University of Aeronautics and Astronautics), Ministry of Education (NJ20240004), Shanghai Oriental Talent Program (QNJY2024007), Fundamental Research Funds for the Central Universities, Science and Technology Commission of Shanghai Municipality (22DZ2229004).